\documentclass[twocolumn,showpacs,preprintnumbers,amsmath,amssymb,prl]{revtex4}
\usepackage{graphicx}
\usepackage{dcolumn}
\usepackage{bm}
\begin{document}
\title{Homogeneity and Size Effects on the Liquid-Gas Coexistence Curve}
\author{Al. H. Raduta$^{1,2}$, Ad. R. Raduta$^{1,2}$}
\affiliation{
  $^1$GSI, D-64220 Darmstadt, Germany\\
  $^2$NIPNE, RO-76900 Bucharest, Romania}
\begin{abstract}
  The effects of (in)homogeneity and size on the phase diagram of
  Lennard-Jones fluids are investigated. It is shown that standard
  multifragmentation scenarios (finite equilibrated systems with
  conserved center of mass position and momentum) are implying a
  strong radial inhomogeneity of the system strongly affecting the
  phase diagram. The homogeneity constraint is therefore necessary for
  finite systems in order to align to the ``meaning'' of infinite
  systems phase diagrams. In this respect, a method which deduces the
  equation of state of homogeneous finite systems from the one
  corresponding to bulk matter is designed. The resultant phase
  diagrams show a strong dependence on the system's size.
\end{abstract}

\maketitle

Phase transitions have been studied from a long time in the limit of
infinite matter. However, there are many physical situations away from
the thermodynamic limit at both microscopic and macroscopic scales.
Such systems are presently of special interest as so far there is
little knowledge about their thermodynamical behavior. There are thus
fundamental questions related to these systems such as whether they do
present phase transitions and, if yes, how can these be identified.
(Recent attempts to extend the standard thermodynamics towards the
``small'' systems can be found in Refs. \cite{gross,bor,chomaz,prl}.)

Highly excited nuclear systems are good ``laboratories'' for
thermodynamics studies. Due to the van der Waals type of the
nucleon-nucleon interaction these systems are supposed \cite{siemens}
to exhibit a liquid-gas phase transition as the classical fluids.  The
connection is however not straight forward due to their (relatively)
small number of constituents and the presence of the (long-range)
Coulomb force. The effect of the Coulomb interaction on the nuclear
liquid-gas phase transitions was studied previously (see e.g. Refs.
\cite{lee-mek,prl}). The effects of other features specific to
multifragmentation such as finite size (for attempts to address the
finitness of the system within Hartree-Fock theories see e.g. Refs.
\cite{jaqaman,satpathy}) and degree of homogeneity are still mostly
unknown.  These aspects are addressed in the present paper.

The classical Lennard-Jones 6-12 (LJ) fluid is used for illustrating
the paper's ideas. While neglecting for the quantum effects acting in
nuclear matter, it has the advantages of generality and tractability.
The LJ potential writes:
\begin{equation}
  v_0(r)=4\epsilon \left[\left(\frac{\sigma}{r}\right)^{12}-
    \left(\frac{\sigma}{r}\right)^6\right]
\end{equation}
Two versions of the above potential are considered herein. The first
is the truncated and shifted (TS) LJ potential: $v(r)=v_0(r)-v_0(r_c)$
when $r<r_c$; $v(r)=0$ when $r \ge r_c$. The second is the truncated
and long range corrected (TLRC) LJ potential: $v(r)=v_0(r)$ when
$r<r_c$; $v(r)=0$ when $r \ge r_c$, corrections being subsequently
included in order to account for the effect of the neglected tail
\footnote{ Contributions from the neglected tail to the system's
potential energy and virial energy term [see eq. (\ref{eq:virial})]
are taken respectively as:\\ $\Delta v=\frac12 \rho A
\int_{r_c}^{\infty} {\rm d}{\bf r}v(r)=8\pi \rho A
\epsilon[\sigma^{12}/(9r_c^9) -\sigma^6/(3r_c^3)]$; $\Delta
\mathcal{V}=\frac12 \rho A \int_{r_c}^{\infty} {\rm d}{\bf
r}(r~\partial v(r)/\partial r) =8\pi \rho A
\epsilon[4\sigma^{12}/(9r_c^9) -2\sigma^6/(3r_c^3)]$.}.  In the
present work we use $r_c=2.5~\sigma$. Phase diagrams of such fluids
are subsequently constructed for various situations.  Phase
transitions in finite systems being here addressed, an adequate
representaion is necessary in order to identify them.  For example,
the $P(V)|_T$ curves corresponding to an isochore canonical ensemble
will prompt the (first order) phase transition through backbendings
{\em even in the case of small systems} \cite{prl,npa_c}. Pressure can
be easily evaluated starting from its canonical definition: $P=T
\partial \ln Z(\beta,V)/\partial V$, where $Z(\beta,V)$ is the
system's canonical partition function:
\begin{equation}
  Z(\beta,V)=\frac1{A!~\lambda_T^{3 A}} 
  \prod_{i=1}^A \int {\rm d}{\bf r}_i 
  \exp \left(-\beta \sum_{i<j}^A v(r_{ij}) \right),
 \label{eq:Z}
\end{equation}
where $\beta\equiv 1/T$ and $\lambda_T$ is the thermal wavelength.
The result is the following
virial expression:
\begin{equation}
  P=\frac{A T}V-\frac1{3V}\left<\sum_{i<j}^A r_{ij} 
    \frac{\partial v(r_{ij})}{\partial r_{ij}}\right>,
  \label{eq:P}
\end{equation}
where $\left<~\right>$ has the meaning of canonical average.  When
conservation of the system's center of mass (c.m.) and c.m.  momentum
is also considered in $Z(\beta,V)$, one simply has to change $A$ into
$A-1$ in eqs. (\ref{eq:Z}) [the power of $\lambda_T$] and (\ref{eq:P})
[the first term]. While rather redundant for very large systems ($A
\rightarrow \infty$), these constraints give important effects when
dealing with small systems. Expressions like (\ref{eq:P}) can be
easily evaluated by means of Metropolis simulations. Knowing the
formal expression of $Z(\beta,V)$, (\ref{eq:Z}), the statistical
simulation of the corresponding canonical ensemble at constant volume
is straight-forward: Pick a randomly chosen particle and make a random
variation of its initial position (in a fixed interval $\Delta V$).
Then, consider this move according to the acceptance $\exp(-\beta
\Delta v_t)$, where $\Delta v_t$ is the change in the system's total
potential energy. The simulation can be easily adapted to the case of
the conserved system's c.m. and c.m. momentum: Two randomly chosen
particles are moved simultaneously with $\Delta {\bf r}$ and $-\Delta
{\bf r}$ such that the system's c.m. remains fixed ($\Delta {\bf r}$
being a variation randomly chosen in the volume interval $\Delta V$).
Then, the move is considered according to the same acceptance.

The present study is started by considering the case of a system at fixed
temperature $T$, composed of $A$ particles interacting via TS LJ
potential, contained into a spherical recipient of volume $V$, having
the c.m. constrained to coincide with the center of the recipient.
The corresponding canonical ensemble is simulated as described
earlier.  Then, isothermal $P(V)$ curves can be evaluated by means of
eq.  (\ref{eq:P}), adapted to the considered conservation laws.
Subsequently, the borders of the liquid-gas coexistence region are
evaluated by performing Maxwell constructions on all $P(V)|_T$ curves
bending backwards. The resulting phase diagrams (in temperature vs.
density representation) corresponding to two systems of different
sizes $A=20$ and $A=50$ are represented in the upper part of Fig. 1.
The increase of the critical temperature with the system's size can be
observed. A striking feature of these phase diagrams is the small
densities corresponding to the borders of the liquid-gas coexistence
regions. In this respect, note that the liquid border is situated at
densities smaller than 0.2 $\sigma^{-3}$ which differs a lot from the
LJ phase diagrams corresponding to infinite homogeneous systems
\cite{hansen-verlet} where the liquid border goes up 0.8
$\sigma^{-3}$. The reason for this discrepancy can be easily
understood from the lower part of Fig. 1. There, the radial density
profiles corresponding to three sample points chosen from the
coexistence line of the A=50 system are represented. In all three cases
the system appears to be inhomogeneous, its density varying from
higher values (towards the center of the recipient) to very small ones
(towards the recipient's walls). The low density tails of these radial
profiles are therefore inducing the above mentioned effect on the
global density of the system (calculated as $A/V$). In particular,
note that at small values of the distance from the recipient's
center the densities of the considered sample points tend to
be consistent with the ones from Ref. \cite{hansen-verlet}. Obviously,
this inhomogeneity effect is dictated by the c.m. conservation
constraint which forces the larger clusters to stay towards the
center of the recipient and the smaller ones towards the borders. This
example is particularly important in the context of multifragmentation
where the system's c.m. is naturally conserved from event to event and
similar density profiles are expected to occur.
\begin{figure}
\vspace*{-.1cm} 
  \hspace*{-0.7cm} 
\includegraphics[height=9cm]{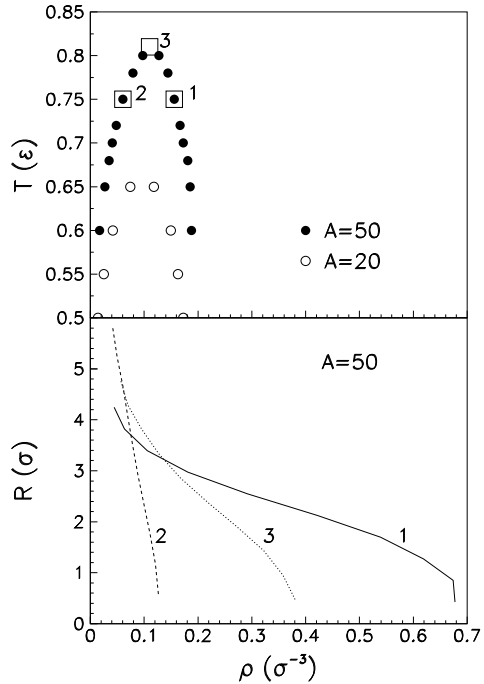}
\caption{Upper panel: Phase diagrams for two systems ($A=20$ and
  $A=50$) with conserved c.m. position and c.m. momentum. Three sample
  points from the $A=50$ phase diagram are represented by
  squares. Lower panel: Radial density profiles corresponding to the
  sample points from the upper part of the figure. The correspondence
  between curves and sample points is given by numbers.}
\label{fig:1}
\end{figure}
\begin{figure}
  \hspace*{-1.5cm} 
    \includegraphics[height=7cm]{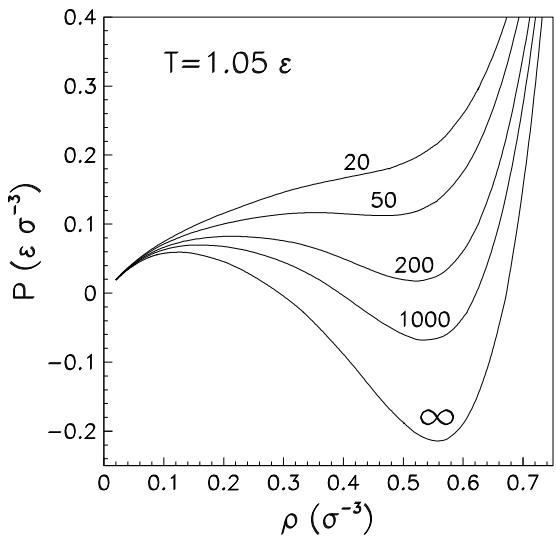}
  \caption{Pressure versus density curves corresponding to
    systems of various sizes ($A$) at fixed temperature, $T=1.05
  \epsilon$, evaluated with eq. (\ref{eq:P1}). The corresponding size
  of the system is specified on each curve.}
  \label{fig:2}
\end{figure}

Therefore, in order to meaningfully compare the phase diagram
calculated for finite systems with the ones corresponding to infinite
systems an extra constraint has to be imposed: {\em the homogeneity}.
The infinite and homogeneous system can be simply approached within
the (above-described) Metropolis simulation by implementing a cubic
recipient with periodic boundary conditions (PBC). (No c.m.
conservation constraint has to be imposed.) A number of 200 particles
interacting via TLRC or TS LJ potentials are placed in a cubic box
with PBC.  The very good agreement between our calculated phase
diagram corresponding to the TLRC case (see Fig. 3, the ``$\infty$''
curve) and the one from Ref.  \cite{hansen-verlet} is confirming the
accuracy of the method here employed. The PBC simulations are used for
evaluating the bulk ``virial energy'' per particle, defined as:
\begin{equation}
  \mathcal{V}=\frac1{3A}\left<\sum_{i<j}^A r_{ij} 
    \frac{\partial v(r_{ij})}{\partial r_{ij}}\right>
  \label{eq:virial}
\end{equation}
Eq. (\ref{eq:P}) can be now translated into:
\begin{equation}
  P=\rho \left(T-\mathcal{V} \right),
  \label{eq:P1}
\end{equation}
where $\rho$ is the system's density. Therefore, in an infinite and homogeneous
system, the virial energy per unity of volume can be written as:
$\tilde{\mathcal{V}}=\rho \mathcal{V}$. 
Given the above definitions one may perform surface corrections for
evaluating the $\mathcal{V}$ term corresponding to finite
systems. Since the system is supposed to be homogeneous, its total virial
energy can be expressed as the difference between a bulk term and a
surface one:
\begin{equation}
  \mathcal{V}_t= \mathcal{V}_b(A-A_s f)=
  \tilde{\mathcal{V}}_b(V-V_s f),  
  \label{eq:vt}
\end{equation}
where the index $b$ specifies that the respective term is a bulk one,
$A_s$ and $V_s$ are respectively the number of particles from the
surface and the ``volume'' of the surface, and finally, $1-f$ is the
ratio between the virial energy of a particle from the surface and the
virial energy of a bulk particle. While for very large values of the
recipient's radius ($R$) $1-f$ is rigorously equal to 1/2 (i.e.  the
particles on the surface have half the number of nearest neighbors
they have in bulk), for small values of $R$ (the case of small
systems) curvature corrections have to be applied to this factor. In
the spirit of the previous definition, $f$ can be fairly approached by
the ratio between the surface of spherical cap situated {\em outside}
the recipient, corresponding to a sphere of radius $\sigma$ having the
center on the surface of the recipient and $4\pi \sigma^2$. After some
algebraic manipulation one gets:
\begin{equation}
  f=\frac12\left(1+\frac{\sigma}{2R}\right).
\end{equation}
Note that when $R\rightarrow\infty$ then $f\rightarrow1/2$.
Considering the surface ``width'' equal to $\sigma$ (i.e. no two
particles from that region, situated on the same radius, attract each
other) one can re-express eq. (\ref{eq:vt}) as:
\begin{equation}
    \mathcal{V}_t=\mathcal{V}_b A \left(1-\frac{V_s}V f\right),
\end{equation}
so that:
\begin{equation}
    \mathcal{V}=\frac{\mathcal{V}_t}A=
    \mathcal{V}_b 
    \left\{ 1- \frac12\left(1+\frac{\sigma}{2R}\right)
    \left[1-\left(1-\frac{\sigma}R\right)^3\right]
  \right\}.
  \label{eq:v}
\end{equation}
Eq. (\ref{eq:v}) can be further expressed in terms of 
$\rho$ using the identity 
$\sigma/R=[4\pi~\rho~\sigma^3 /(3 A)]^{1/3}$. 
Subsequently, eq. (\ref{eq:P1}) 
[with $\mathcal{V}$ given by eq. (\ref{eq:v})]
can be applied for calculating the pressure of a finite system of size
$A$ at various values of $\rho$.

An illustration of the results of the method is given in Fig. 2 where
$P(\rho)|_T$ curves, corresponding to the temperature
$T=1.05~\epsilon$ are represented for various sizes of the system.
One can observe that the depth of the backbending of the curve is
diminishing as the size of the source is decreasing such that at
system sizes as small as $A=20$ the backbending completely disappears.
\begin{figure}
  \hspace*{-1.5cm} 
    \includegraphics[height=7cm]{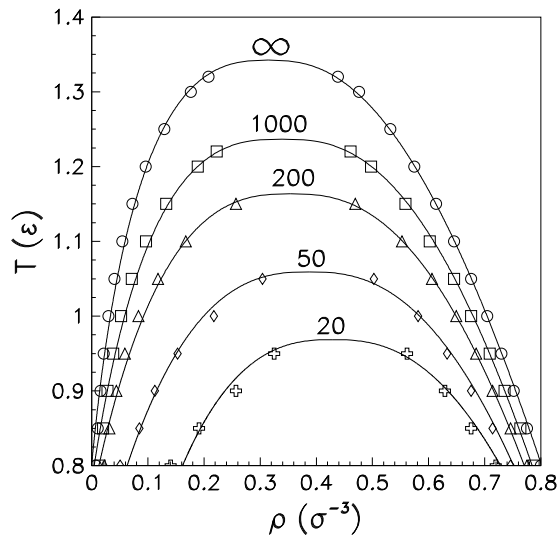}
  \caption{Phase diagrams corresponding to the TLRC LJ fluid,
    corresponding to different sizes of the system. The system's size
  is specified on top of each diagram. Points calculated
  via Maxwell constructions are represented with symbols. Full
  lines are fits of eq. (\ref{eq:ggnhm}) on the 
  calculated points.}
  \label{fig:3}
\end{figure}
  
Maxwell constructions have been further performed on the $P(V)|_T$
curves corresponding to various values of $T$ and $A$ allowing thus
the construction of phase diagrams for various sizes of the system.
The results corresponding to the TLRC and TS LJ forms of potential are
given in Figs. 3 and 4 respectively. The points from the coexistence
region borders obtained via Maxwell constructions can be further
interpolated in order to get an estimation of the critical point and a
clearer view on the systems' phase diagrams by means of the Guggenheim
scaling relations for the coexistence curve:
\begin{equation}
  \rho_{\pm}=\rho_c\left(1+a~\epsilon \pm b~\epsilon^{\beta}\right),
  \label{eq:ggnhm}
\end{equation}
where $\rho_+$ corresponds to the liquid branch, $\rho_-$ to the vapor
branch of the coexistence curve, $\rho_c$ is the critical density,
$\epsilon\equiv(T_c-T)/T_c$ ($T_c$ being the critical temperature) and
$\beta$ is the critical exponent of the coexistence curve. Eq.
(\ref{eq:ggnhm}) is fitted on the points obtained via the Maxwell
construction method by adjusting the parameters: $T_c$, $\rho_c$,
$\beta$, $a$ and $b$. As observed in Figs. 3 and 4 the quality of the
obtained fits is very good.  (It is worth mentioning that for the TS
potential, infinite system case, $\beta=0.34$ which sharply
corresponds to the liquid-gas universality class.) For both considered
potentials the critical temperature is drastically decreasing and the
critical density is increasing with decreasing the size of the source.
In particular, note that important deviations from the ``$\infty$''
phase diagram can be observed even for the phase diagram of a system
as large as $A=10^4$ (see Fig. 4). This result is particularly
important since it proves that thermodynamically the bulk limit is not
(even by far) attainable in nuclear multifragmentation experiments
(where the equilibrated systems formed are usually smaller then
$A=300$).
\begin{figure}
  \hspace*{-1.5cm} 
    \includegraphics[height=7cm]{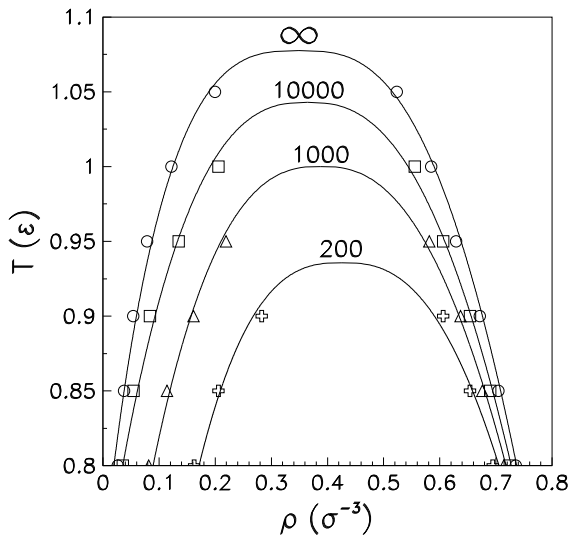}
  \caption{Same as Fig. 3 but for a TS LJ fluid.}
  \label{fig:4}
\end{figure}

Summarizing, homogeneity and size effects on the system's phase
diagram have been discussed in the framework of the classical LJ
fluid.  Phase diagrams are constructed using a method adequate for
revealing the coexistence region even for small systems: Maxwell
constructions on canonical $P(V)|_T$ curves.  It was shown that
standard scenarios like finite systems with conserved c.m., physically
consistent with the multifragmentation phenomenon imply a strong
radial inhomogeneity in the systems importantly affecting their phase
diagrams. Thus, in order to align to the {\em meaning} of the phase
diagrams corresponding to the infinite fluids \cite{ggnhm}, an extra
constraint has to be introduced for the case of the finite systems:
{\em homogeneity}.  To this aim, a method for constructing the
$P(V)|_T$ curves corresponding to homogeneous finite systems based on
making surface corrections on the bulk virial energy term was
designed. This method makes possible the deduction of the $P(V)|_T$
curves corresponding to any size of the system, $A$, by using the
information embedded in the bulk $P(V)|_T$ curves. The resulting phase
diagrams corresponding to homogeneous systems of various sizes show a
strong dependence on the size of the system. In this respect, the
critical temperature is drastically decreasing when the size of the
system is reduced. For example, important deviations from the bulk
phase diagram can be noticed even for systems as large as $A=10^4$.
This means that the largest equilibrated systems formed in nuclear
multifragmentation experiments are not even close to the bulk limit.
To this effect one should, of course, couple the inhomogeneity
(discussed earlier) and the Coulomb ones. And as shown in Ref.
\cite{prl} Coulomb is independently bringing a very important
contribution towards lowering the system's critical point.
\begin{acknowledgments}
This work was supported by the Alexander von Humboldt Foundation.
\end{acknowledgments}
 

\end{document}